\documentclass[man, floatsintext]{apa7}

\usepackage[american]{babel}
\usepackage[style=apa,sortcites=true,sorting=nyt,backend=biber]{biblatex}
\usepackage{biblatex}
\usepackage{tabularx}
\usepackage{booktabs}
\usepackage{xcolor}
\usepackage[]{minted} 
\usepackage{subcaption}
\usepackage{amssymb}
\usepackage{amsmath}
\usepackage{float}
\usepackage[justification=centering]{caption}
\usepackage{caption}
\usepackage{multirow}
\usepackage{csquotes}
\usepackage{orcidlink}
\usepackage[document]{ragged2e}
\usepackage{hyperref}
\hypersetup{colorlinks=true}
\definecolor{LightGray}{gray}{0.95}
\setminted[python]{frame=lines,framesep=2mm,baselinestretch=1.2}

\DeclareLanguageMapping{american}{american-apa} 
\title{StanBKT: Rethinking Parameter Estimation in Bayesian Knowledge Tracing} 
\shorttitle{StanBKT}

\authorsnames[1,2,3,4,4,2]{
Siddhartha Pradhan\orcidlink{0009-0004-9977-1442},
Yanping Pei\orcidlink{0009-0005-1628-3177},
Morgan Lee\orcidlink{0000-0002-9839-9608},
Puyuan Zhang\orcidlink{0000-0003-3217-2870},
Erin Ottmar\orcidlink{0000-0002-9487-7967},
Adam C. Sales\orcidlink{0000-0003-0416-0610}
}
\authorsaffiliations{
{Department of Data Science},
{Department of Mathematical Sciences},
{Department of Computer Science},
{Department of Social Science and Policy Studies}
\\
{Worcester Polytechnic Institute,  Worcester, MA, USA}
}

\abstract{Bayesian Knowledge Tracing (BKT) remains one of the most widely used and interpretable student modeling approaches in intelligent tutoring systems and educational data mining. However, most existing BKT implementations rely on expectation-maximization or related optimization procedures that provide only point estimates, limiting uncertainty quantification and principled statistical comparison across learners and conditions. To address these limitations, we introduce StanBKT, an open-source Python package for estimating BKT models using Bayesian inference in Stan. StanBKT provides a unified framework supporting multiple inference methods, including Hamiltonian Monte Carlo (HMC), variational inference, Pathfinder, and optimization-based estimation, while preserving the interpretability and hidden Markov structure of classical BKT. The package supports standard, grouped, and hierarchical BKT formulations, flexible prior specification, posterior predictive inference, and downstream analysis utilities for visualization and diagnostics. We evaluate StanBKT on both large-scale observational and controlled experimental educational datasets. Using the ASSISTments 2020 dataset, we demonstrate that the inference methods supported in StanBKT achieve comparable predictive performance while differing in their trade-offs between computational efficiency and posterior fidelity. We further demonstrate how posterior inference enables principled comparison of condition-specific BKT parameters in an educational intervention study examining perceptual cue manipulations. Results illustrate how uncertainty quantification facilitates more reliable interpretation of differences in learning, forgetting, guessing, and slipping parameters across experimental conditions. Overall, StanBKT extends BKT beyond point estimation by providing an accessible and flexible framework for probabilistic student modeling, uncertainty quantification, and hierarchical inference in educational data mining research.}

\keywords{Bayesian Knowledge Tracing, student modeling, uncertainty quantification, Bayesian estimation, hierarchical BKT}

\authornote{
\raggedright
\paragraph{Corresponding Author} 
Siddhartha Pradhan, Worcester Polytechnic Institute, 100 Institute Road, Worcester, MA 01609. Email: sppradhan@wpi.edu.

\paragraph{Acknowledgments}
We gratefully acknowledge the Stan Development Team for their continued development and maintenance of the Stan probabilistic programming language, which provides the core inference engine underlying StanBKT. This material is based upon work supported by the National Science Foundation under Grant [2300764] awarded to Worcester Polytechnic Institute.
}

\begin{document}
\maketitle

\section{Introduction}

Over the past three decades, intelligent tutoring systems (ITS) have emerged through interdisciplinary collaboration across education, computer science, cognitive science, and educational psychology \parencite{andersonCognitiveTutorsLessons1995, heffernanASSISTmentsEcosystemBuilding2014, wangExaminingApplicationsIntelligent2023}. In contrast to traditional classroom settings, where student understanding is typically assessed through periodic exams or assignments, ITS enable continuous assessment during learning by maintaining computational representations of student knowledge and misconceptions, commonly referred to as student models \parencite{bakerEnsemblingPredictionsStudent2011, siemensLearningAnalyticsEducational2012}. These models support adaptive instruction by estimating a learner’s evolving mastery of underlying skills in real time.

Bayesian Knowledge Tracing (BKT) is one of the most widely studied and applied student modeling approaches \parencite{bakerMoreAccurateStudent2008, nedungadiPredictingStudentsPerformance2015, pardosKTIDEMIntroducingItem2011, takamiEvaluatingEffectivenessBayesian2024, xuBayesianbasedKnowledgeTracing2023, zambranoInvestigatingAlgorithmicBias2024}. It estimates a student’s probability of mastering a given skill (or knowledge component) based on their sequence of observed responses over time. Despite its conceptual simplicity, BKT has been extensively used in both research and deployed tutoring systems due to its interpretability and ability to capture learning dynamics from interaction data.

Beyond prediction, BKT has also been widely used to support adaptive decision-making in educational systems. Estimating student mastery enables tutoring platforms to deliver personalized interventions targeting specific knowledge gaps. Prior work has demonstrated that BKT-based student models have been incorporated into a wide range of intelligent tutoring applications. These include adaptive language tutoring systems that select instructional content based on estimated knowledge states \parencite{schoddeAdaptiveRobotLanguage2017}, explanation-generation systems that leverage slip and guess parameters to personalize feedback \parencite{takamiEducationalExplainableRecommender2021}, and hybrid approaches that use deep learning models, such as Transformers, to estimate BKT parameters while preserving the interpretability of the underlying student model and achieving strong predictive performance \parencite{badrinathOptimizingBayesianKnowledge2025}. Collectively, these systems demonstrate how BKT can function as a core inference component within larger educational technologies that support real-time adaptation and personalized learning.

More recently, researchers have also used BKT as an interpretability tool to study how instructional interventions and learning behaviors influence student learning dynamics. As BKT parameters (i.e., learning, forgetting, guessing, and slipping) correspond to interpretable cognitive processes, they provide a structured way to quantify how different educational conditions affect both short-term performance and long-term learning. For instance, \textcite{beckDoesHelpHelp2008} used BKT to analyze the impact of help interventions in a reading ITS, finding that help increased learning rates while also influencing immediate performance through changes in guessing behavior.

Similarly, recent work by \parencite{zhangUsingBayesianKnowledge} applied BKT to an open-source arithmetic learning dataset to examine how perceptual cue manipulations (e.g., color and spacing congruency) affect student learning. By analyzing changes in guess and slip parameters across conditions, they identified instructional conditions that improved immediate performance and others that enhanced long-term learning, consistent with theories of desirable difficulty \parencite{bjorkDesirableDifficultiesTheory2020, koedingerItBetterGive}. Importantly, these parameter-level findings aligned with observed behavioral patterns, demonstrating the potential of BKT as a tool for validating cognitive and educational theories using empirical data.

As data-driven models become increasingly integrated into educational research and practice, there is a growing need for accessible and transparent tools that can be used by researchers beyond the educational data mining and machine learning communities. Due to its interpretable hidden Markov structure and psychologically meaningful parameters, Bayesian Knowledge Tracing (BKT) is particularly well suited for this role, enabling broader adoption across the learning sciences and related disciplines.

To support this goal, several open-source implementations and platforms have been developed to lower the barrier to entry for BKT modeling, including software libraries such as \texttt{pyBKT} \parencite{badrinath_pybkt_2021,bulutIntroductionBayesianKnowledge2023} and systems such as OATutor \parencite{pardosOATutorOpensourceAdaptive2023}. However, most existing implementations rely on expectation-maximization or related optimization procedures that produce only point estimates, limiting researchers' ability to quantify uncertainty and make principled comparisons across conditions. These limitations motivate the development of more flexible and unified frameworks for probabilistic student modeling. In response, we introduce StanBKT, a Python package built on Stan that enables full Bayesian inference for BKT models, supporting prior specification, hierarchical extensions, and posterior-based analyses while preserving the interpretability of the classical BKT framework.

\section{Model Description} \label{sec:model-description} 
Standard (or classical) Bayesian Knowledge Tracing \parencite{corbettKnowledgeTracingModeling1994} is a probabilistic generative model and can be viewed as a special case of a Hidden Markov Model (HMM) in which both the latent state and observed outcomes are dichotomous. The observed outcome $Y_t \in \{0,1\}$ indicates whether the student correctly answers item $P_t$. The latent state at time $t$, denoted $L_t$, represents whether a student has mastered a given skill $k$ while working through a sequence of items $P_{1:t}$. Throughout this paper, we use $P(L_t)$ as shorthand for the probability of mastery at time $t$, that is, $P(L_t = 1)$.

\begin{figure}[h]
    \centering
    \includegraphics[width=0.7\textwidth]{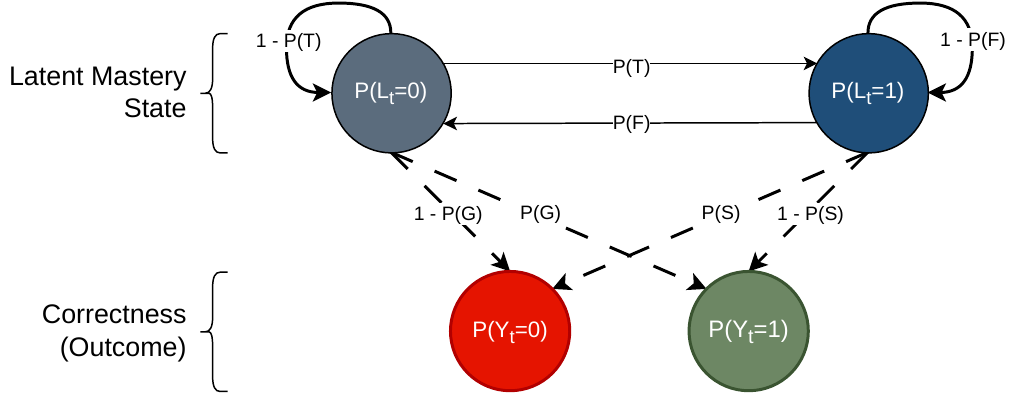}
    \caption{State transition diagram illustrating latent knowledge states and emission probabilities in BKT}
    \label{fig:bkt-diagram}
\end{figure}

The model assumes that knowledge evolves as a first-order Markov chain. The transition dynamics are governed by $P(T)$, the probability of transitioning from the unlearned to the learned state, and, in formulations that allow forgetting, $P(F)$, the probability of transitioning from the learned to the unlearned state. Additionally, the observed response $Y_t$ is generated from the latent state $L_t$ through an emission process governed by the slip probability $P(S)$, which captures the chance of an incorrect response given mastery, and the guess probability $P(G)$, which captures the chance of a correct response given non-mastery. The overall model structure and these dependencies are illustrated in Figure~\ref{fig:bkt-diagram}.

Formally, the BKT model is parameterized by $\boldsymbol{\theta} = \{P(T), P(F), P(S), P(G), \pi_{\text{know}}\}$, corresponding to the learning probability (learn rate), forgetting probability (forget rate), slip probability, guess probability, and initial probability of mastery\footnote{In most BKT literature this is typically referred to as the prior, however, to avoid confusion with Bayesian prior distributions, we refer to this as the initial probability of mastery or initial state distribution.}. Given these parameters $\boldsymbol{\theta}$, the model defines a joint distribution over latent knowledge states $L_{1:T}$ and observed responses $Y_{1:T}$ (i.e., complete data likelihood):
\begin{equation}
\label{eq:bkt-joint}
P(L_{1:T}, Y_{1:T} \mid \boldsymbol{\theta})
=
P(L_1)
\prod_{t=2}^{T} P(L_t \mid L_{t-1}, \boldsymbol{\theta})
\prod_{t=1}^{T} P(Y_t \mid L_t, \boldsymbol{\theta}).
\end{equation}

This factorization encodes the standard first-order Hidden Markov Model assumptions. In particular, the current latent state depends only on the previous latent state \big(i.e., $P(L_t \mid L_1, \dots, L_{t-1}) = P(L_t \mid L_{t-1})$\big), and each observation is conditionally independent of all other variables given the corresponding latent state \big(i.e., $P(Y_t \mid Y_1, \dots, Y_T, L_1, \dots, L_T) = P(Y_t \mid L_t)$\big). Under these assumptions, the transition distribution $P(L_t \mid L_{t-1})$ in Equation~\ref{eq:bkt-joint} is specified by the learn probability $P(L_t = 1 \mid L_{t-1} = 0)$ and the forget probability $P(L_t = 0 \mid L_{t-1} = 1)$. The emission distribution $P(Y_t \mid L_t)$ captures how observed responses arise from the latent knowledge state and is parameterized by the slip probability \big($P(S) = P(Y_t=0 \mid L_t=1)$\big) and the guess probability \big($P(G) = P(Y_t=1 \mid L_t=0)$\big).

Given a trained BKT model, the latent mastery state is continuously updated during online inference. Specifically, after observing a new response at time $t$, the latent state is updated via Bayes' rule. First, the conditional probability $P(L_t \mid Y_{t})$ is computed depending on whether the student learned and applies the skill correctly or incorrectly. For a correct response ($Y_t = 1$):
\[
P(L_t \mid Y_{t} = 1)
=
\frac{P(L_t)\, \big(1 - P(S)\big)}
{P(L_t)\, \big(1 - P(S)\big) + \big(1 - P(L_t)\big)\, P(G)},
\]
or an incorrect response ($Y_t = 0$):
\[
P(L_t \mid Y_t = 0)
=
\frac{P(L_t)\, P(S)}
{P(L_t)\, P(S) + \big(1 - P(L_t)\big)\, \big(1-P(G)\big)}.
\]
Next, this conditional probability is used to update the probability of mastery of the next time step $L_{t+1}$:
\[
P(L_{t+1})
=
P(L_t \mid Y_{t})\,\big(1 - P(F)\big) + \big(1 - P(L_t \mid Y_t)\big)\,P(T).
\]
Given these components, the model induces a one-step-ahead predictive (1-SAP) distribution for the next observation, which is obtained by marginalizing over the posterior predictive distribution of the latent state at time $t+1$
\[
P(Y_{t+1} = 1)
= 
P(L_{t+1})\,\big(1-P(S)\big) + \big(1 - P(L_{t+1})\big)\,P(G).
\]

While the preceding equations describe how the model is used for online inference and prediction at each time step, parameter estimation requires evaluating how well the model explains the entire observed response sequence $Y_{1:T}$. The likelihood is obtained by integrating out the latent mastery states from the joint factorization in Equation~\ref{eq:bkt-joint}:
\[
P(Y_{1:T} \mid \boldsymbol{\theta})
= \sum_{L_{1:T}} P(L_{1:T}, Y_{1:T} \mid \boldsymbol{\theta}),
\]
which is typically evaluated using the forward algorithm, a dynamic programming procedure for efficient computation. In prior BKT packages, such as \texttt{PyBKT} by \textcite{badrinath_pybkt_2021} and its \texttt{R} counterpart, \texttt{BKT} by \textcite{yuan_bkt_2026}, the parameters $\boldsymbol{\theta}$ are estimated using the Expectation--Maximization (EM) algorithm, specifically the Baum--Welch algorithm \parencite{yang_statistical_2015}.  

\subsection{Bayesian Formulation}

Unlike the frequentist formulation of BKT, which typically relies on EM for point estimation, the Bayesian approach treats the model parameters as random variables and places prior distributions over them. The objective in this formulation is to infer the posterior distribution $P(\boldsymbol{\theta} \mid Y_{1:T})$, which explicitly captures parameter uncertainty and allows it to propagate to downstream analyses such as mastery trajectories, posterior predictive checks, and comparisons across intervention groups.

In this framework, priors are specified for each component of $\boldsymbol{\boldsymbol{\theta}}$. For example:
\[
P(T) \sim \text{Beta}(\alpha_T, \beta_T), \quad
P(F) \sim \text{Beta}(\alpha_F, \beta_F),
\]
and analogous priors for $P(S)$, $P(G)$, and $\pi_{\text{know}}$. Subsequently, the joint distribution factorizes as:
\[
P(\boldsymbol{\theta}, L_{1:T}, Y_{1:T})
=
P(\boldsymbol{\theta})\, P(L_1 \mid \boldsymbol{\theta})\prod_{t=2}^{T} P(L_t \mid L_{t-1}, \boldsymbol{\theta})\prod_{t=1}^{T} P(Y_t \mid L_t, \boldsymbol{\theta}).
\] Given observed responses $Y_{1:T}$, Bayesian inference targets the posterior distribution
\[
P(\boldsymbol{\theta}, L_{1:T} \mid Y_{1:T})
=
\frac{
P(\boldsymbol{\theta})\, P(L_{1:T}, Y_{1:T} \mid \boldsymbol{\theta})
}{
P(Y_{1:T})
}
\propto
P(\boldsymbol{\theta})\, P(L_{1:T}, Y_{1:T} \mid \boldsymbol{\theta}),
\]
where the marginal likelihood (i.e., evidence) is obtained by integrating over parameters and summing over latent state sequences:
\[
P(Y_{1:T})
=
\int \sum_{L_{1:T}}
P(\boldsymbol{\theta})\, P(L_{1:T}, Y_{1:T} \mid \boldsymbol{\theta})\, d\boldsymbol{\theta}.
\]

However, exact inference is intractable due to the summation over exponentially many latent state sequences and the continuous integration over $\boldsymbol{\theta}$. In practice, the posterior is approximated using either Markov chain Monte Carlo (MCMC) or variational inference (VI). Both approaches rely on repeated evaluation of the complete data likelihood, $P(L_{1:T}, Y_{1:T} \mid \boldsymbol{\theta})$, as defined in Equation~\ref{eq:bkt-joint}.

\subsection{BKT Variants}

Several works have extended the classical BKT model to account for variation across learners, items, and instructional contexts. As shown in Table~\ref{tab:bkt_variants}, these extensions primarily differ in how they relax assumptions about fixed parameters, such as independence across items, static vs contextual slip and guess rates, or homogeneous learner behavior. Additional modifications include individualizing parameters to the student, adding additional mastery states, or modeling partial correctness \parencite{saric-grgicTwentyfiveYearsBayesian2024}. 

Several works have also explored deep learning-based variants to further enhance the performance of knowledge tracing models for mastery and correctness predictions. One of the first such approaches is Deep Knowledge Tracing, which uses recurrent neural networks to model student learning over time \parencite{piech2015deep}. Subsequent work has introduced attention-based models, such as SAKT, which use self-attention mechanisms to identify relevant past interactions when predicting future performance \parencite{pandeySelfAttentiveModelKnowledge2019}. More recently, hybrid approaches such as BKTransformer have combined neural architectures with BKT structure by learning data-driven parameterizations of the classical BKT model \parencite{badrinathOptimizingBayesianKnowledge2025}.

Despite these advances, the improved accuracy of deep learning-based knowledge tracing models over classical BKT has been partly attributed to their high-dimensional latent representations and their ability to capture interleaved skill interactions within a single unified model \parencite{monteroDoesDeepKnowledge2018}. Subsequent comparative studies have shown that this performance gap narrows across several datasets when the classical BKT model is extended with the forgetting parameter  \parencite{khajahHowDeepKnowledge2016b, badrinath_pybkt_2021}. Additionally, in comparison to classical BKT and its variants, which assume binary knowledge and outcome nodes, the high-dimensional deep learning-based approaches are uninterpretable and offer little insight into the learning process itself \parencite{saric-grgicTwentyfiveYearsBayesian2024}.

\begin{table}[!htbp]
\begin{threeparttable}
\caption{Some Notable Developments in Knowledge Tracing} \label{tab:bkt_variants}
\centering
\small
\begin{tabular}{p{0.3\textwidth} p{0.6\textwidth}}
\hline
Method & Description \\
\hline
\multicolumn{2}{c}{\textit{Classical Extensions}} \\
\hline
BKT+Forget \parencite{beckDoesHelpHelp2008,yudelsonMultifactorApproachStudent2008} & Extends BKT by introducing a non-zero forgetting probability, allowing mastery to decay over time. \\
\hline
Contextual Guess and Slip \parencite{bakerMoreAccurateStudent2008} & Estimates guess and slip parameters using contextual features rather than assuming they are fixed. \\
\hline
Help Model \parencite{beckDoesHelpHelp2008} & Incorporates tutor interventions as an additional factor influencing student performance. \\
\hline
KT-PPS \parencite{pardosModelingIndividualizationBayesian2010} & Conditions prior knowledge estimates on early student performance (e.g., first-item correctness). \\
\hline
KT-IDEM \parencite{pardosKTIDEMIntroducingItem2011} & Extends BKT by making guess and slip parameters item-specific. \\
\hline
\multicolumn{2}{c}{\textit{Deep Learning Based Methods}} \\
\hline
DKT \parencite{piech2015deep} & One of the first approaches to use recurrent neural networks for modeling student learning over time. \\
\hline
SAKT \parencite{pandeySelfAttentiveModelKnowledge2019} & Uses a self-attention mechanism to identify relevant past interactions for predicting future performance. \\
\hline
BKTransformer \parencite{badrinathOptimizingBayesianKnowledge2025} & Uses a neural network-based parameter generation approach to improve BKT flexibility and performance. \\
\hline
\end{tabular}
\end{threeparttable}
\end{table}

\section{The StanBKT Package}
While Bayesian Knowledge Tracing and its extensions have been widely adopted, most existing implementations rely on point-estimate methods such as EM, with none supporting full Bayesian inference. Although researchers can use general-purpose probabilistic programming frameworks such as Stan for full Bayesian inference, it requires substantial modeling effort and expertise. This ultimately limits the accessibility of using Bayesian inference for BKT parameter estimation in educational data mining and learning analytics research.

To address this gap, we introduce StanBKT, a Python implementation of Bayesian Knowledge Tracing built on Stan. StanBKT provides a unified interface for model specification and inference, combining the interpretability of classical BKT with the flexibility and uncertainty quantification of Bayesian methods.

\subsection{Motivation}

Compared with recent developments in knowledge tracing, particularly deep learning-based approaches, Bayesian Knowledge Tracing remains attractive for its interpretable latent-state structure and computationally efficient hidden Markov formulation. However, most existing implementations rely on expectation-maximization or related optimization procedures, which yield only point estimates. This is limiting in practice, as uncertainty over learning, guessing, slipping, and initial knowledge is often of substantive methodological interest.

Bayesian estimation is, therefore, a natural alternative to BKT for several reasons. First, it provides principled uncertainty quantification over all model parameters and latent states. The uncertainty is also propagated to downstream analysis, such as correctness estimation. Second, it allows the incorporation of prior knowledge through prior distributions, enabling regularization and the encoding of domain and researchers' assumptions. Third, Bayesian inference can help address well-known identifiability issues in BKT, where multiple parameter values may explain the same observed data equally well. In such cases, informative or well-structured priors can reduce model degeneracy and improve stability. Finally, Bayesian formulations naturally extend to hierarchical models, enabling partial pooling across students, skills, or items, which is particularly beneficial in sparse-data settings common in educational experiments.

Despite these advantages, most existing implementations do not support full Bayesian inference in a flexible and accessible way. StanBKT was developed to address this limitation while preserving the interpretability of classical BKT. The package uses Stan as a backend for model fitting and provides a high-level Python interface for model specification, parameter estimation, and extraction of model-based quantities of interest. Overall, StanBKT offers a flexible estimation framework supporting both Bayesian inference and optimization-based fitting.

\subsection{Software Overview} \label{sec:software-overview}
StanBKT is an open-source Python package for estimating classical BKT models and several extensions using Bayesian inference in Stan. The package is designed to be both flexible and accessible, enabling users to specify, fit, and analyze BKT models with minimal setup and prior experience.

The source code is publicly available on GitHub at \href{https://github.com/SiddharthaPradhan/StanBKT}{github.com/SiddharthaPradhan/StanBKT}. StanBKT supports Windows, macOS, and Linux, and can be installed via \texttt{pip} using \texttt{pip install stanbkt}. Additionally, as Stan models are transpiled to \texttt{C++} and compiled before execution, a compatible \texttt{C++} toolchain is required. Platform-specific installation instructions, along with comprehensive user documentation and worked examples, are provided in the online documentation at \href{https://stanbkt.readthedocs.io}{stanbkt.readthedocs.io}.

\begin{figure}
    \centering
    \includegraphics[width=0.6\linewidth]{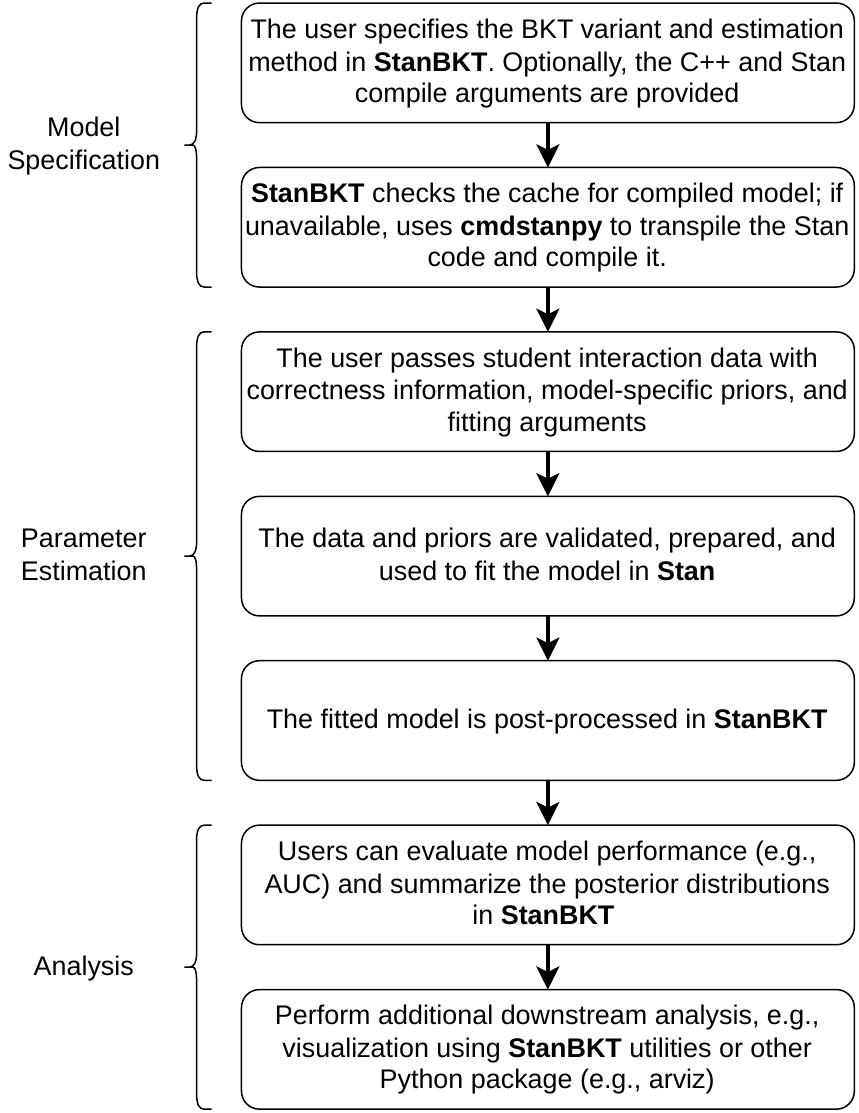}
    \caption{A high-level overview of the model fitting workflow used in StanBKT.}
    \label{fig:stanbkt-flow}
\end{figure}

Models are specified and fitted in StanBKT using the workflow shown in Figure~\ref{fig:stanbkt-flow}. In the model specification stage, the user first specifies a BKT variant (e.g., Standard BKT) and an estimation method (e.g., MCMC), along with optional \texttt{C++} or \texttt{Stan} compilation arguments such as \texttt{STAN\_THREADS} for multi-threading. Given these inputs, StanBKT attempts to retrieve a matching precompiled \texttt{C++} model from cache. If no cached version is available, StanBKT uses \texttt{cmdstanpy} to translate the \texttt{Stan} model into \texttt{C++} and compile it into executable machine code. This compilation step is required for all \texttt{Stan} models and is handled automatically within StanBKT.

In the parameter estimation stage, the user passes student interaction data, including correctness information, along with model-specific priors and additional fitting arguments to the model’s \texttt{fit()} method. The data is expected to be in long format, consistent with prior packages and logging conventions of standard intelligent tutoring systems. The inputs are then validated and prepared (e.g., checking required columns and formats), after which the model is fitted in \texttt{Stan} using the specified inference procedure. Once fitting is complete, the resulting model is post-processed within StanBKT to facilitate downstream use and to standardize access to model outputs.

In the analysis stage, users can evaluate model performance using standard metrics such as AUC and summarize posterior distributions of the parameters. This includes generating quantities such as posterior means and credible intervals, as well as extracting model-based predictions. Additional downstream analysis, including visualization of posterior predictions or learning trajectories, can be performed using built-in StanBKT utilities or external Python packages such as \texttt{arviz} \parencite{Martin2026}.

\subsection{Parameter Estimation}
As described in the \nameref{sec:software-overview} section, StanBKT utilizes \texttt{Stan} as the inference backend. Consequently, all inference algorithms implemented in \texttt{Stan} are available for parameter estimation. By default, \texttt{Stan} employs Hamiltonian Monte Carlo (HMC) \parencite{neal2011mcmc}, specifically the No-U-Turn Sampler (NUTS) \parencite{hoffmanNoUTurnSamplerAdaptively2011}, an adaptive variant of HMC. Compared to traditional gradient-free samplers such as random-walk Metropolis, HMC-based methods leverage gradient information to efficiently explore the posterior distribution. This results in lower autocorrelation between samples, improved effective sample sizes, and faster mixing, particularly in moderate- to high-dimensional parameter spaces. Additionally, HMC does not require conjugate priors, enabling flexible model specification \parencite{carpenterStanProbabilisticProgramming2017}.

However, these advantages come at a substantial computational cost. Each iteration of HMC requires multiple evaluations of the model likelihood and its gradients, making it significantly more expensive than optimization-based approaches such as EM. As a result, while HMC provides asymptotically exact posterior inference and enables uncertainty quantification, it is often orders of magnitude slower in practice.

\begin{table}[h!]
\begin{threeparttable}    
    \caption{Supported Parameter Estimation Methods in StanBKT and Suggested Use Cases}
    \label{tab:estimation-methods}
    \centering
    \begin{tabular}{p{0.3\textwidth} p{0.6\textwidth}}
        \hline
        Estimation Method & Use Case \\
        \hline
        Markov Chain Monte Carlo (NUTS) & Full Bayesian inference providing asymptotically exact posterior samples. Recommended when uncertainty quantification, or rigorous posterior analysis, is required, at the cost of high computational expense. \\
        \hline
        Variational Inference (VI) & Approximate Bayesian inference via optimization of the ELBO. Suitable for large datasets or exploratory analysis where approximate posterior summaries are sufficient, but may underestimate uncertainty and simplify parameter dependencies. \\
        \hline
        Pathfinder &  Provides a fast Gaussian or mixture approximation to the posterior, often offering improved posterior accuracy over variational inference while remaining substantially faster than MCMC. Can also be used to initialize HMC. \\
        \hline
        Maximum Likelihood Estimation (MLE) / Maximum a Posteriori (MAP)  & Point estimation obtained by maximizing the likelihood (MLE) or posterior (MAP) using gradient-based optimization. This does not quantify parameter uncertainty and collapses the posterior to a single mode. Useful for initial exploration and assessing model fit. \\
        \hline
    \end{tabular}
\end{threeparttable}
\end{table}

To address this trade-off, \texttt{Stan} provides faster \textit{approximate} inference methods that sacrifice posterior accuracy in exchange for computational efficiency. A summary of available inference methods and their typical use cases is provided in Table~\ref{tab:estimation-methods}.

Variational inference (VI) in \texttt{Stan} is implemented through Automatic Differentiation Variational Inference (ADVI) \parencite{kucukelbirAutomaticDifferentiationVariational2017}, which approximates the posterior $P(\boldsymbol{\theta} \mid Y_{1:T})$ with a tractable distribution $q(\boldsymbol{\theta})$ by maximizing the evidence lower bound (ELBO). Similarly, Pathfinder \parencite{zhang2022pathfinder} provides an alternative optimization-based approximation that constructs a Gaussian (or mixture) approximation to the posterior. Compared to standard VI, Pathfinder more effectively captures parameter correlations, often yielding better approximations while remaining significantly faster than HMC.

These approaches are substantially faster than MCMC and scale well to large datasets. However, the resulting approximation is biased and often underestimates posterior uncertainty and simplifies dependencies among parameters. Yet these methods provide practical alternatives when full MCMC is computationally infeasible. In particular, Pathfinder can be used to obtain a high-quality initial approximation that can be used to initilize HMC, thereby reducing warmup (burn-in) time.

\subsection{Model Variants} \label{sec:model-variants}

StanBKT supports multiple BKT formulations that differ in how parameters are shared across students and groups. These variants range from full pooling to partial pooling, allowing the model to adapt to different data structures and experimental designs. StanBKT organizes these formulations into three main types based on the underlying pooling assumption:

\begin{enumerate}
    \item \textbf{Standard BKT}: The classical formulation in which a single parameter set 
    $\boldsymbol{\theta} = \{P(T), P(F), P(S), P(G), \pi_{\text{know}}\}$ 
    is shared across all students and interactions for a given skill. This corresponds to a fully pooled model and assumes homogeneous learning and response behavior across students and items.

    \item \textbf{Multi BKT}: Extends the standard formulation by allowing parameters to vary across a predefined grouping structure (e.g., classrooms, treatment conditions, problems, cohorts, or individual students). Let $j \in \{1, \dots, N_j\}$ index grouping structure. For each skill, a separate parameter set $\boldsymbol{\theta}_j$ is estimated for each group. This removes the pooling assumption and enables direct modeling and comparison of parameters across different groups.

    \item \textbf{Hierarchical BKT}: Introduces partial pooling through hierarchical priors on group-specific parameters. Instead of estimating each $\boldsymbol{\theta}_j$ independently, parameters are decomposed into a global component and group-level deviations. For each parameter $\phi \in \boldsymbol{\theta}$, we define
    \[ 
    \text{logit}({\phi_j}) = \text{logit}(\boldsymbol{\phi}) + \delta_{\phi_j}, \quad \delta_{\phi_j} \sim \mathcal{N}(0, \sigma^2_\phi), 
    \]
    
    where $\phi_0$ is the population-level parameter and $\delta_{\phi_j}$ captures group-specific variation. The variance $\sigma^2_\phi$ is learned from data and controls the degree of shrinkage: smaller values induce stronger pooling toward the global parameter, while larger values allow greater between-group variability. This formulation stabilizes and regularizes estimates in sparse data contexts while preserving systematic differences across groups.
\end{enumerate}

These model classes can also be flexibly configured within StanBKT. For example, users may start from the Standard BKT specification and disable components such as forgetting by fixing $P(F) = 0$.\footnote{While fixing parameters is supported in StanBKT, a fully Bayesian alternative is to place a highly concentrated prior on the parameter.} Similarly, any parameter can be fixed to a user-specified value, in which case it is treated as known and inference is conditioned on it.

Across all model types, the initial mastery parameter $\pi_{\text{know}}$ can optionally be modeled at the student level. For the Standard BKT model, this corresponds to the KT-PPS formulation. StanBKT supports two approaches to estimate individualized $\pi_{\text{know}}$. In the first, $\pi_{\text{know}}$ is estimated directly as a free parameter from interaction data. In the second, we implement a novel estimation strategy, where $\pi_{\text{know}}$ is modeled as a function of student covariates:
\[
\text{logit}(\pi_{\text{know}_j}) = x_j^\top \beta,
\]
where $x_j$ denotes covariates for student $j$ and $\beta$ are the corresponding coefficients. This formulation links initial mastery to observed student characteristics via a linear predictor on the logit scale, while allowing systematic variation across students to be explicitly captured.


\subsection{Prior Specification}
\label{sec:prior-specification}
Priors\footnote{Here, ``priors'' refers to prior distributions in the Bayesian sense, rather than the initial probability of knowing a skill, which is also commonly called a “prior” in the BKT literature.} in StanBKT are specified on a transformed (logit) scale to ensure that all probability parameters remain constrained to the unit interval while allowing flexible prior control in an unconstrained real-valued space. Specifically, each probability parameter in the BKT model, including the learning rate $P(T)$, forgetting rate $P(F)$, slip $P(S)$, guess $P(G)$, and initial knowledge probability $\pi_{\text{know}}$, is mapped through a logistic transformation of an underlying Gaussian prior. This parameterization ensures unconstrained sampling during inference while preserving probabilistic validity after transformation. 

For a generic probability parameter $\phi \in (0,1)$, the model assumes:
\[
\text{logit}(\phi) \sim \mathcal{N}(\mu_\phi, \sigma_\phi^2),
\quad \text{where} \quad
\phi = \text{logit}^{-1}(z) = \frac{1}{1 + \exp(-z)}.
\]

Unlike the learn and forget parameters, the guess $P(G)$, and slip $P(S)$ parameters are constrained to the interval $[0, 0.5]$. While informative priors may encourage reasonable posterior estimates for $P(G)$ and $P(S)$, they may not guarantee avoidance of degenerate solutions or sufficiently mitigate inherent identifiability issues in the model. To avoid such issues, these parameters are explicitly restricted using hard bounds, preventing the model from assigning unrealistically high probabilities of guessing or slipping. In practice, this constraint is implemented by parameterizing $P(G)$ and $P(S)$ on a scaled (half) logit scale. Specifically, an unconstrained logit parameter $z \in \mathbb{R}$ is mapped to the probability space via an inverse-logit (i.e., sigmoid) transformation followed by scaling, i.e., $P(G), P(S) = 0.5 \cdot \text{logit}^{-1}(z)$. For example, if a prior $z \sim \mathcal{N}(0, \sigma^2)$ is specified for the logit-scale parameter of $P(G)$ or $P(S)$, the resulting distribution on the probability scale is centered at $0.25$, with its spread controlled by $\sigma^2$.

Extending beyond the core BKT parameters, StanBKT also supports additional prior specifications for components introduced in more flexible model variants. For example, in the hierarchical BKT variant, users may define a Bayesian prior to specify the between-subject variability $ \sigma_\phi^2$. A complete overview of these priors and their default configuration is provided in the online documentation (c.f. LINK HERE).

Users may also choose not to specify explicit priors. In \texttt{Stan}, this corresponds to an implicit uniform prior over the parameter space. In such cases, the posterior is driven primarily by the likelihood. Under this setting, the maximum a posteriori (MAP) estimate is equivalent to the maximum likelihood estimate (MLE), since the prior contributes a constant factor over the valid parameter space.


\subsection{Posterior Inference and Prediction}
\label{sec:posterior-inf-and-pred}

After model fitting, StanBKT provides utilities for latent-state estimation, prediction, and visualization. Given a fitted BKT model with parameters $\boldsymbol{\theta}$, one-step-ahead prediction estimates the probability of a correct response at the next opportunity, $P(Y_{t+1} \mid Y_{1:t})$, which requires estimating the latent mastery state $P(L_{t+1} \mid Y_{1:t})$. These filtered estimates are commonly used in online or adaptive learning settings, where predictions must be updated sequentially as new responses are observed.

In addition to filtered estimates, StanBKT supports smoothed latent-state estimation, which conditions on the full response trajectory $Y_{1:T}$. Specifically, smoothing estimates $P(L_{t+1} \mid Y_{1:T})$, allowing retrospective inference of student mastery states using both past and future observations. Such estimates are useful for post-hoc analysis and interpretation of student learning trajectories.

For estimation methods that produce posterior draws (e.g., MCMC, variational inference, or particle filtering), predictions may be generated separately for each posterior sample $\boldsymbol{\theta^{(i)}
}$, where $i \in \{1, \dots, N_{\text{draws}}\}$. This yields a posterior predictive distribution, $P(Y_{t+1}^{(i)} \mid Y_{1:t}, \boldsymbol{\theta^{(i)}})$. Alternatively, predictions may be computed using a posterior point estimate of the parameters (e.g., the posterior mean or median), yielding a single predictive estimate, $P(Y_{t+1} \mid Y_{1:t}, \boldsymbol{\hat{\theta}})$.

Posterior predictive distributions allow uncertainty in the model parameters to propagate into downstream analyses and visualizations, whereas point estimates provide a computationally efficient summary suitable for rapid prototyping and evaluation.

\subsection{Worked Example} \label{subsec:worked-example}

To illustrate how StanBKT is used in practice, we provide a complete worked example that walks through the core workflow for fitting and analyzing a Bayesian Knowledge Tracing model. This example is intended to familiarize readers with the package interface and demonstrate how model specification, parameter estimation, and posterior analysis are integrated within a single reproducible framework.

We begin by simulating data from a standard BKT model and then fit the same model using StanBKT. The workflow follows Figure~\ref{fig:stanbkt-flow} and consists of three stages: model specification, parameter estimation, and analysis.

\subsubsection{Data Generation}
\label{subsubsec:data-gen}
In this example, we will generate data using utilities provided by StanBKT. The general data-generation procedure follows the standard HMM framework described in \nameref{sec:model-description}. For the classical BKT formulation, given the parameters $\boldsymbol{\theta} = \{P(T), P(F), P(G), P(S), \pi_{\text{know}}\}$, the initial mastery state $L_0^{(j)}$ is sampled as $L_0^{(j)} \sim \text{Bernoulli}(\pi_{\text{know}})$ for each student $j \in \{1, \dots, N_j\}$. 

Next, for each timepoint $t \in \{1, \dots, T\}$, we simulate the transition dynamics for mastery $L_t^{(j)} \rightarrow L_{t+1}^{(j)}$:
\[
P(L_{t+1}^{(j)} = 1 \mid L_t^{(j)}) =
\begin{cases}
P(T)     & \text{if } L_t^{(j)} = 0 \\
1 - P(F) & \text{if } L_t^{(j)} = 1 
\end{cases}\] The observed responses $Y_{1:T}^{(j)}$ is then generated conditional on the latent mastery state:
\[
P(Y_t^{(j)} = 1 \mid L_t^{(j)}) =
\begin{cases}
P(G)     & \text{if } L_t^{(j)} = 0 \\
1 - P(S) & \text{if } L_t^{(j)} = 1
\end{cases}
\]

In this example, we generate data for $N_j = 100$ students working on a single skill with $N_p = 30$ problems. We then randomly sample 80\% of the generated interactions without replacement to simulate missing or skipped problems. The first five rows of the resulting dataset are shown in Table~\ref{tab:student_problem_data}. The table includes all required columns for fitting the standard BKT model.

\begin{listing}[H]
\begin{minted}{python}
from stanbkt.utils import sim_simple_BKT
# define true param values
bkt_params = {
    "prior": 0.4,
    "learn": 0.04,
    "forget": 0.01,
    "guess": 0.1,
    "slip": 0.05,
}
data_df = sim_simple_BKT(
    n_students=100,
    n_problems=30,
    n_kcs=1,
    frac=0.8,  # sample 80%
    rng_seed=1234,  # random seed
    **bkt_params,  # true param values
)
\end{minted}
\caption{Simulate interaction data for Classical BKT using StanBKT utilities}
\label{listing:simulation}
\end{listing}

\begin{table}[H]
\centering
\begin{threeparttable}
\caption{The first few rows from the generated data}
\label{tab:student_problem_data}
\begin{tabular}{lllll}
\toprule
\texttt{student\_id} & \texttt{problem\_id} & \texttt{correct} & \texttt{timestamp} & \texttt{kc\_id} \\
\midrule
stu\_0 & prob\_0 & 0 & 2024-01-01 00:00:00 & kc\_0 \\
stu\_0 & prob\_1 & 0 & 2024-01-01 00:01:00 & kc\_0 \\
stu\_0 & prob\_2 & 0 & 2024-01-01 00:02:00 & kc\_0 \\
stu\_0 & prob\_4 & 0 & 2024-01-01 00:04:00 & kc\_0 \\
stu\_0 & prob\_5 & 0 & 2024-01-01 00:05:00 & kc\_0 \\
\bottomrule
\end{tabular}
\tablenote{Correct is coded as 1 = correct response, 0 = incorrect response. KC = knowledge component.}
\end{threeparttable}
\end{table}

\subsubsection{Model Specification}
For estimation, we use MCMC, which relies on Hamiltonian Monte Carlo (HMC) with the NUTS sampler. In addition, we set `\texttt{STAN\_THREAD: True}', enabling compilation of the model with multi-threading support.

\begin{listing}[H]
\begin{minted}{python}
from stanbkt.models import StandardBKT
from stanbkt.fits import FitMethod
from stanbkt.utils import VerbosityLevel
model = StandardBKT(
    fit_method=FitMethod.MCMC,
    verbose=VerbosityLevel.WARN,
    cpp_compile_kwargs={"STAN_THREADS": True},
)
\end{minted}
\caption{Specify a BKT model with the preferred estimation method}
\label{listing:model-specification}
\end{listing}

\subsubsection{Parameter Estimation}
After specifying the model, StanBKT requires three components to estimate the parameters: interaction data containing student correctness, prior distributions corresponding to the selected model, and additional arguments specific to the fit method (e.g., the number of warmup and sampling iterations for MCMC).

For the interaction data, StanBKT requires columns representing student identifiers, problem identifiers, correctness, sequential ordering, and (for grouped models) a group identifier. To process this data, a column mapping must be provided to align dataset column names with those expected by StanBKT.

\begin{listing}[H]
\begin{minted}{python}
from stanbkt.utils import ColumnNames
col_mapping = {
    ColumnNames.STUDENT_ID: "student_id",
    ColumnNames.PROBLEM_ID: "problem_id",
    ColumnNames.KC_ID: "kc_id",
    ColumnNames.CORRECTNESS: "correct",
    ColumnNames.ORDER: "timestamp",
}
\end{minted}
\caption{Specifying the column name mapping}
\label{listing:col-mapping}
\end{listing}

To encode prior beliefs about the parameters, we use the appropriate priors class (e.g., \texttt{StandardPriors} for \texttt{StandardBKT}). In this example, we assume minimal prior knowledge about parameter values and therefore adopt weakly informative priors. Specifically, we set $logit({\pi_{\text{know}}}) \sim \mathcal{N}(0,2)$, which corresponds to a mean of 0.5 on the probability scale with 95\% intervals from 0.019 to 0.980.

The same interpretation does not directly apply to the guess and slip parameters. As described in \nameref{sec:prior-specification}, these parameters are constrained to $[0, 0.5]$ on the probability scale. Their priors are defined on a ``half-scaled-logit'' scale, meaning values are scaled by 0.5 after transformation. Consequently, $logit\big({P(G)}\big) \sim \mathcal{N}(0,2)$ corresponds to a mean of 0.25 with 95\% intervals from 0.009 to 0.490 on the probability scale. All other parameters without priors specified will use default values.

\begin{listing}[H]
\begin{minted}{python}
from stanbkt.models import StandardPriors
priors = StandardPriors(
    # pi_know
    pi_know_mu=0,
    pi_know_std=2,
    # guess
    guess_mu=0,
    guess_std=2,
)
\end{minted}
\caption{Specify priors for specific parameters on the logit scale}
\label{listing:priors}
\end{listing}

Finally, we specify MCMC-specific fitting arguments. We use 500 warmup iterations and draw 500 posterior samples for each of the four chains. The chains are run in parallel, with two threads assigned per chain to improve parallelization and reduce computation time. With all components defined, the model can be fit to the data.

\begin{listing}[H]
\begin{minted}{python}
from stanbkt.fits import MCMCFitOptions
fit_opts = MCMCFitOptions(
    seed=1234,  # seed
    iter_warmup=500,
    iter_sampling=500,
    chains=4,
    parallel_chains=4,
    threads_per_chain=2,
)
model.fit(data_df, 
            stan_fit_options=fit_opts, 
            priors=priors)
\end{minted}
\caption{Fit the model to the data with selected priors and fit options}
\label{listing:model-fitting}
\end{listing}

\subsubsection{Parameter Summaries and Diagnostics}

For models that return posterior draws (i.e., MCMC, VB, and PF), parameter estimates can be inspected using \mintinline{python}|model.summary()|. This function returns standard posterior summary statistics, including the mean, median, and standard deviation. For MCMC-based inference, additional diagnostics are available, such as the effective sample size (ESS), Monte Carlo standard error (MCSE), and the potential scale reduction factor ($\hat{R}$), which is commonly used to assess chain convergence. The summary results for this example are provided in Table~\ref{tab:posterior_summary}.

\begin{table}[H]
\centering
\begin{threeparttable}
\caption{Posterior Summary of the Fitted Model}
\label{tab:posterior_summary}
\begin{tabular}{lccccc}
\toprule
Parameter & Mean & MCSE & SD & Median & $\hat{R}$ \\
\midrule
$\pi_{\text{know}}$ & 0.452 & 0.001 & 0.055 & 0.450 & 1.002 \\
$P(T)$ & 0.047 & 0.000 & 0.007 & 0.046 & 1.002 \\
$P(F)$ & 0.022 & 0.000 & 0.005 & 0.022 & 1.002 \\
$P(G)$ & 0.091 & 0.000 & 0.011 & 0.091 & 1.002 \\
$P(S)$ & 0.036 & 0.000 & 0.006 & 0.036 & 1.001 \\
\bottomrule
\end{tabular}
\tablenote{MCSE = Monte Carlo standard error; SD = standard deviation; $\hat{R}$ = Gelman--Rubin convergence statistic. Additional columns are hidden for brevity. The true data-generating parameter values were: $\pi_{\text{know}} = 0.40$, $P(T) = 0.04$, $P(F) = 0.01$, $P(G) = 0.10$, and $P(S) = 0.05$.
}
\end{threeparttable}
\end{table}

\begin{figure}[ht!]
\centering

\begin{subfigure}{0.8\linewidth}
    \centering
    \includegraphics[width=\linewidth]{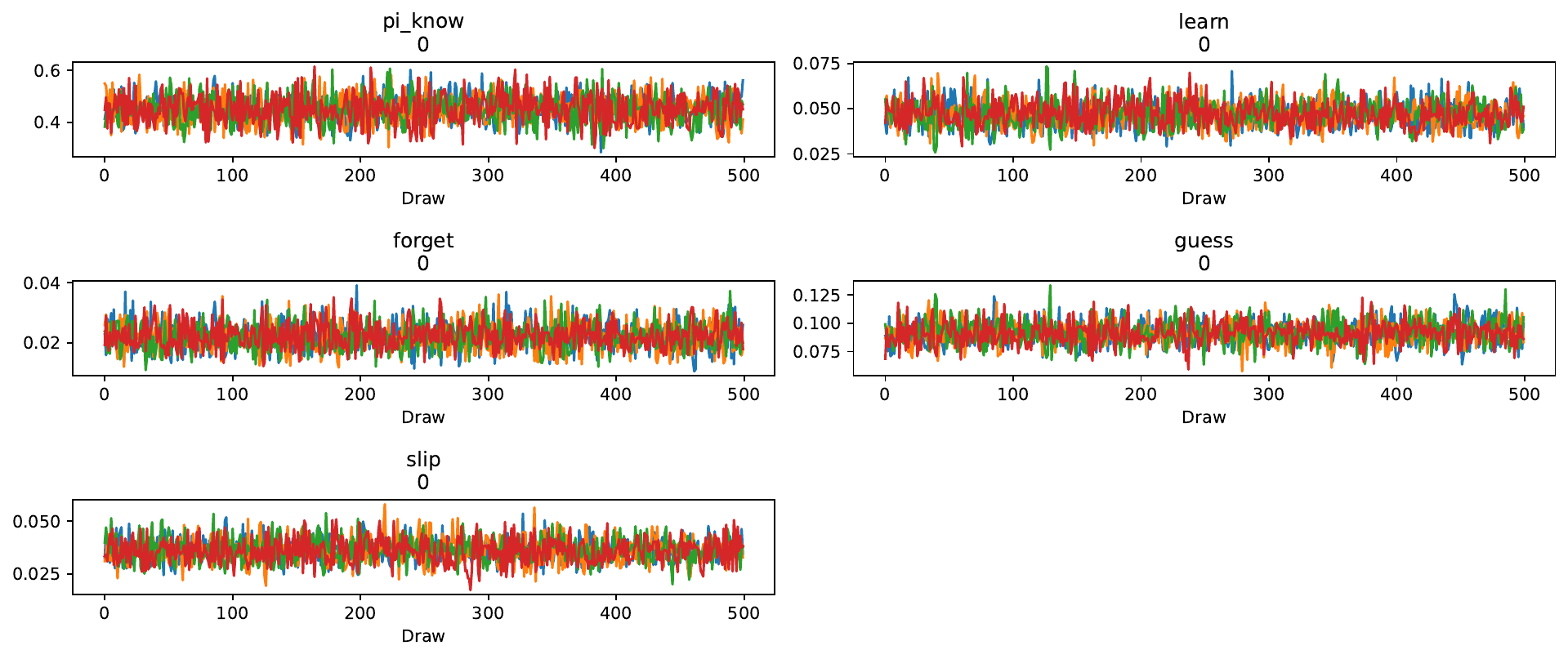}
    \caption{Trace plots for the parameters}
    \label{fig:param-trace}
\end{subfigure}

\begin{subfigure}{0.8\linewidth}
    \centering
    \includegraphics[width=\linewidth]{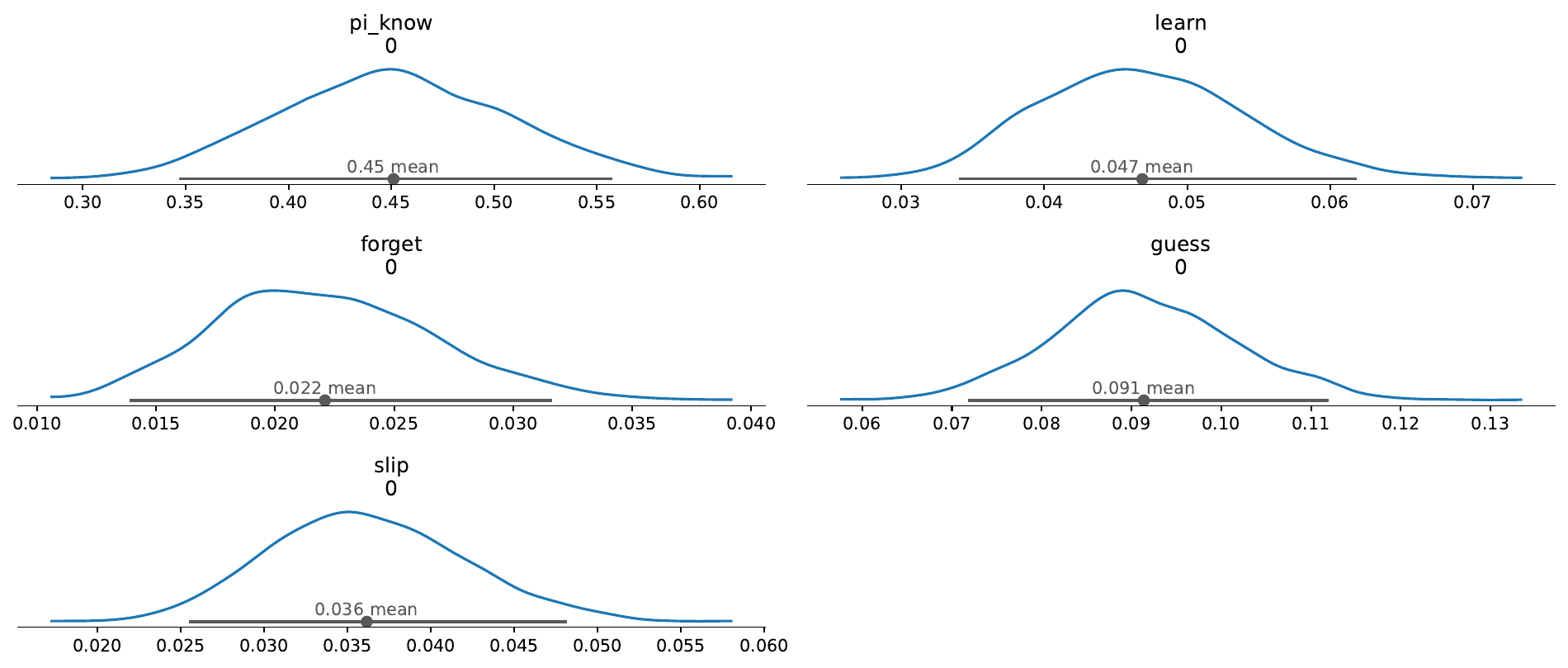}
    \caption{Posterior distribution plots for the parameters}
    \label{fig:param-dist}
\end{subfigure}

\caption{Plotting the posterior for the parameters in the worked example}
\label{fig:example-panels}
\end{figure}

StanBKT also provides utilities for visualizing posterior parameter distributions and MCMC trace plots, allowing users to inspect parameter uncertainty, chain mixing, and convergence behavior across sampling iterations. Listing~\ref{listing:param-plots} demonstrates how to generate these visualizations for a fitted model. Figure~\ref{fig:param-trace} displays the trace plots for each parameter across MCMC iterations, which can be used to assess mixing and convergence. Figure~\ref{fig:param-dist} shows the corresponding posterior distributions with 95\% credible intervals, providing a summary of parameter uncertainty.

\begin{listing}[H]
\begin{minted}{python}
from stanbkt.plot import plot_dist, plot_trace
kc0_fit = model.fits.get_fit("kc_0")
trace_fig = plot_trace(
    kc0_fit,
    params=["pi_know", "learn", 
            "forget", "guess", "slip"],
    col_wrap=2,
)
trace_dist = plot_dist(
    kc0_fit,
    params=["pi_know", "learn", 
            "forget", "guess", "slip"],
    ci_prob=0.95,
    col_wrap=2,
)
\end{minted}
\caption{Trace and distribution plots for the parameters.}
\label{listing:param-plots}
\end{listing}

\subsubsection{Analysis Using Fitted Model}
\label{sec:worked-pred-example}

As described in the \nameref{sec:posterior-inf-and-pred} section, StanBKT supports prediction using either full posterior draws or posterior point estimates of the model parameters. When posterior draws are available for the parameters, predictions can be generated separately for each sampled parameter set $\theta^{(i)}$. Calling \mintinline{python}{model.predict_posterior()} yields posterior predictive distributions that preserve parameter uncertainty. Alternatively, for computationally efficient evaluation, \mintinline{python}{model.predict()} may be used to generate predictions using a posterior point estimate of the parameters, such as the posterior mean (i.e.,  $\mathbb{E}[\theta]$).

Additionally, latent mastery state estimates and correctness predictions may be computed using either filtered (i.e., \mintinline{python}{model.predict()} and \mintinline{python}{model.predict_posterior()}) or smoothed (\mintinline{python}{model.predict_smoothed()}) and \mintinline{python}{model.predict_smoothed_posterior()}) inference. Filtered estimates condition only on past observations, whereas smoothed estimates condition on the full response trajectory. The appropriate approach depends on the intended application. For example, filtered inference is typically used in online adaptive learning settings, while smoothed inference is often more appropriate for retrospective analyses and visualization.

In this example, we focus on one-step-ahead online prediction using filtered inference. Specifically, we generate posterior predictions for the latent mastery state, $P(L_t \mid Y_{1:t-1})$, and the probability of a correct response, $P(Y_t \mid Y_{1:t-1})$. The result of executing the code in Listing~\ref{listing:posterior-summary} is given in Table~\ref{tab:posterior_prediction_summary}.

\begin{listing}[H]
\begin{minted}[breaklines]{python}
from stanbkt.utils import posterior_summary
predictions_draws = model.predict_posterior_draws(
        data_df,
        column_mapping=col_mapping,
    ) 
# summarize the draws
posterior_summary(predictions_draws)
\end{minted}
\caption{Generating and summarizing posterior predictions.}
\label{listing:posterior-summary}
\end{listing}

\begin{table}[!htbp]
\centering
\begin{threeparttable}
    
\caption{Posterior Summaries for Predicted Latent Knowledge and Correctness}
\label{tab:posterior_prediction_summary}
\begin{tabular}{llllcccc}
\toprule
& & & &
\multicolumn{2}{c}{$P(L_t)$} &
\multicolumn{2}{c}{$P(C_t)$} \\
\cmidrule(lr){5-6}
\cmidrule(lr){7-8}
KC ID & Student ID & Problem ID & Correct &
Mean & SD &
Mean & SD \\
\midrule
kc\_0 & stu\_0 & prob\_0 & 0 & 0.451 & 0.058 & 0.485 & 0.050 \\
kc\_0 & stu\_0 & prob\_1 & 0 & 0.077 & 0.011 & 0.159 & 0.013 \\
kc\_0 & stu\_0 & prob\_2 & 0 & 0.050 & 0.008 & 0.135 & 0.011 \\
kc\_0 & stu\_0 & prob\_4 & 0 & 0.049 & 0.008 & 0.134 & 0.011 \\
kc\_0 & stu\_0 & prob\_5 & 0 & 0.049 & 0.008 & 0.134 & 0.011 \\
\bottomrule
\end{tabular}
\tablenote{$P(L_t)$ denotes the posterior probability that the student has learned (i.e., mastered) the knowledge component at opportunity $t$, and $P(C_t)$ denotes the posterior probability of a correct response. Additional posterior summaries, including medians and credible intervals, are omitted for brevity.
}
\end{threeparttable}
\end{table}

\begin{figure}
    \centering
    \includegraphics[width=0.8\linewidth]{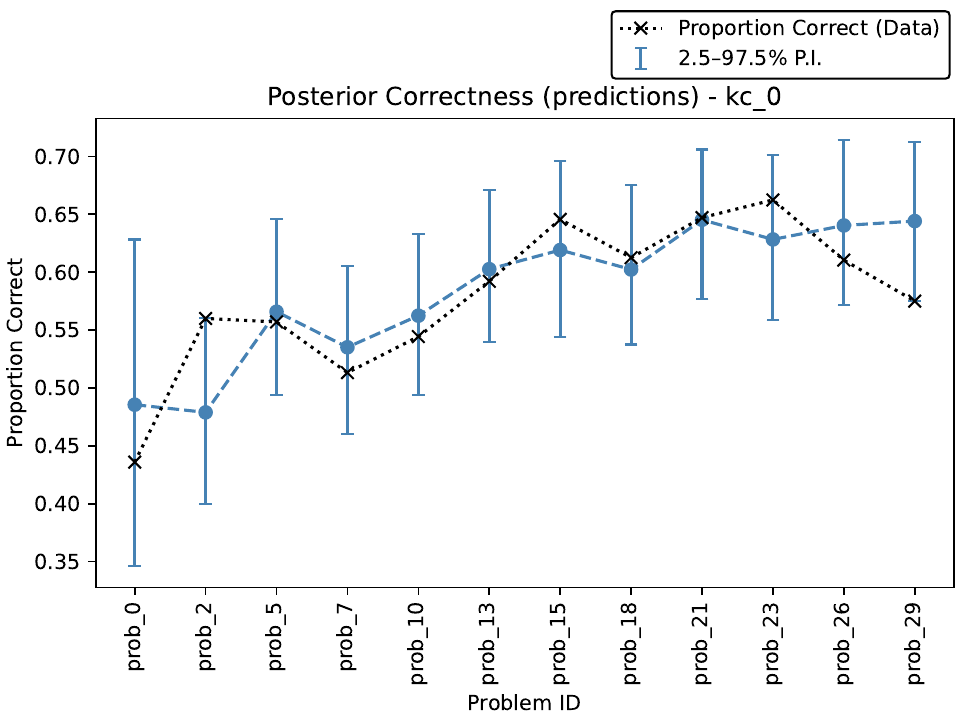}
    \caption{Visualizing true and estimated proportion correct across problems}
    \label{fig:prop-correct}
\end{figure}

Subsequently, posterior correctness predictions can be visualized to assess how well the model captures overall response trends in the data. In this example, we compare the observed proportion correct with the posterior predictive estimates across problems for a selected knowledge component. The resulting figure is presented in Figure~\ref{fig:prop-correct}

\begin{listing}[H]
\begin{minted}{python}
from stanbkt.plot import plot_posterior_correctness
plot_posterior_correctness(
    posterior_pred_kc=predictions_draws["kc_0"],
    data=data_df,
    kc="kc_0",
    type="preds",
    trajectory=True,
    frac=0.4,
)
\end{minted}
\end{listing}

\section{Empirical Evaluation} \label{sec:evaluation}
This section evaluates the practical performance of StanBKT across large-scale and experimental educational datasets. We focus on two facets: (1) how well different inference algorithms scale to realistic datasets while maintaining predictive accuracy, and (2) how the resulting posterior estimates can be used to support comparisons of model parameters across experimental conditions. Together, these evaluations demonstrate both the computational flexibility and inferential capacity of StanBKT.

\subsection{Scalability Evaluation on ASSISTments 2020} \label{subsec:assist-eval}
We assess the scalability and predictive performance of StanBKT on the ASSISTments 2020 dataset, a large-scale educational dataset containing student interaction sequences across multiple skills \parencite{priharStudentEngagementRemote}. The goal is to assess how different inference methods trade off computational efficiency and predictive performance on a realistic, large-scale educational dataset. The dataset is available through OSF at \href{https://osf.io/q7zc5}{osf.io/q7zc5}, and contains the entirety of the interactions of students and teachers within the ASSISTments platform during the 2019-2020 school year. 

For this analysis, we selected the 10 knowledge components (KCs) with the most interactions and removed student sequences with fewer than 10 observations. This subset was chosen because the full dataset contains several KCs with relatively few observations, resulting in highly sparse response sequences that can lead to unstable parameter estimates and unreliable performance comparisons. Restricting the analysis to the most frequently observed KCs ensures that each model is fit on sufficiently informative data while still preserving a realistic large-scale benchmark. The resulting dataset contains 794{,}158 training interactions and 189{,}082 test interactions. The split was made at the student level, such that all interactions from a given student appear in only one split. This setup evaluates the model's ability to generalize to previously unseen students.

Predictive performance was evaluated using the following metrics: 

\begin{itemize}
    \item \textbf{Accuracy}: The proportion of correctly classified responses after thresholding predicted probabilities at 0.5.
    \item \textbf{Area Under the Receiver Operating Characteristic Curve (AUC)}: Measures the model's ability to discriminate between correct and incorrect responses across all possible classification thresholds.
    \item \textbf{Root Mean Squared Error (RMSE)}: Quantifies the average squared deviation between the predicted correctness probabilities and the observed outcomes.
\end{itemize}

Together, these metrics capture complementary aspects of model performance, including classification accuracy, ranking ability, and the overall quality of the predicted probabilities.

We compare several inference methods available in StanBKT, including MCMC, variational inference, Pathfinder, and maximum a posteriori (MAP) estimation, against the \texttt{pyBKT} package, which uses expectation-maximization (EM). All models were fit using the standard BKT model with the forgetting parameter. The comparison focuses on both fit time and predictive performance. Although runtime comparisons are not strictly normalized across methods (e.g., MCMC generates thousands of posterior draws across multiple chains), the reported times provide a practical indication of the computational cost associated with each approach under representative default settings.

All experiments were conducted on a single compute node with 40 allocated CPU cores from dual AMD EPYC 9654 processors and 64~GB of RAM. For MCMC, we used four parallel chains with 1{,}000 warmup iterations and 1{,}000 post-warmup sampling iterations per chain. All other inference methods were run using Stan’s default settings as specified in the Stan Reference Manual \parencite{stan_reference_manual}.

Table~\ref{tab:fit-eval-summary} summarizes fit time and predictive performance across methods using point-estimate predictions. Predictive performance was nearly identical across all methods, with accuracy, AUC, and RMSE differing by only a few decimal places. The primary distinction lies in computational cost. MCMC provides full posterior inference but requires substantially more computation, whereas variational inference, Pathfinder, and MAP offer considerably faster alternatives while maintaining equivalent predictive performance. Among the StanBKT-based methods, MAP was the fastest, requiring only slightly more fitting time than \texttt{pyBKT}, while achieving identical predictive performance to the other inference methods.

\begin{table}[!htbp]
\centering
\begin{threeparttable}
    
\caption{Fit time and test-set predictive performance for pyBKT and StanBKT inference methods using point-estimate predictions.}
\label{tab:fit-eval-summary}

\begin{tabularx}{0.8\linewidth}{X r r r r}
\toprule
Method & Fit Time (s) & Accuracy & AUC & RMSE \\
\midrule
pyBKT (EM) & 3.35 & 0.694 & 0.711 & 0.449 \\
\midrule
\multicolumn{5}{c}{\textit{StanBKT}} \\
\midrule
MCMC & 822 & 0.694 & 0.711 & 0.449 \\
Pathfinder & 151 & 0.694 & 0.711 & 0.449 \\
Variational & 26.8 & 0.694 & 0.711 & 0.449 \\
MAP & 6.83 & 0.694 & 0.711 & 0.449 \\
\bottomrule
\end{tabularx}

\tablenote{All StanBKT methods use posterior point estimates for prediction. For Bayesian methods, point estimates correspond to posterior means. pyBKT results are reported using its default settings.
}
\end{threeparttable}
\end{table}

\subsection{Analysis of Perceptual Cue Effects using Multi BKT} \label{subsec:cue-eval}

While the analysis on the ASSISTments 2020 dataset focused on scalability and predictive performance in a large observational dataset, this second evaluation highlights the inference capabilities of StanBKT in a controlled experimental setting. Specifically, we use the \texttt{MultiBKT} model to estimate condition-specific learning parameters and quantify uncertainty in differences across experimental groups. This example illustrates how StanBKT can facilitate hypothesis testing and parameter comparison in educational experiments.

The experimental dataset is drawn from an open-source perceptual cue intervention study available on Open Science Framework (OSF) at \href{https://osf.io/d3tzq}{osf.io/d3tzq} \parencite{ottmar_data_2025}. The original study included 1{,}110 sixth-grade students from the United States. Following the preprocessing procedure conducted by \textcite{ottmar_data_2025}, we restricted the analysis to the 688 students who completed all required tasks and practice items. These students solved 29 order-of-operations problems, yielding 19{,}952 total problem-solving observations.

\begin{figure}
    \centering
    \includegraphics[width=0.5\linewidth]{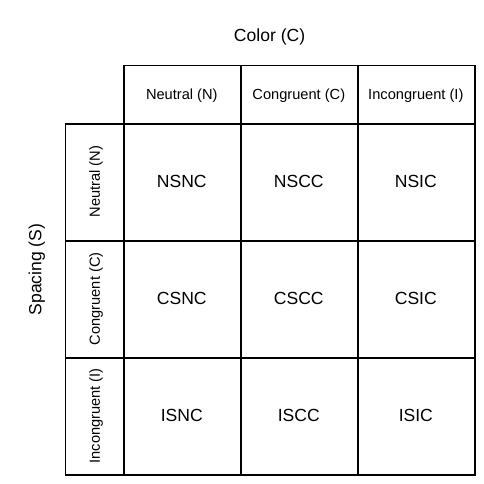}
    \caption{The nine conditions in the $3 \times 3$ factorial design crossing spacing and color cues. Cell labels denote the corresponding spacing and color combinations (e.g., NSCC = Neutral Spacing, Congruent Color).}
    \label{fig:condition-design}
\end{figure}

Participants were randomly assigned to one of nine between-subject conditions defined by a $3 \times 3$ factorial design crossing two perceptual cue manipulations: spacing and color. Each factor included three levels: congruent, incongruent, and neutral, resulting in approximately 70--80 students per condition. Figure~\ref{fig:condition-design} summarizes the experimental design. Additional details regarding the intervention, participant demographics, and study procedures are provided by \textcite{ottmar_data_2025}.

Given that students' responses are conditioned on the perceptual cue they receive during the intervention, researchers may wish to quantify how different interventions affect the latent learning dynamics represented by the BKT parameters. To achieve this, we use the framework presented by \textcite{beckDoesHelpHelp2008}, which introduced the Bayesian Evaluation and Assessment framework. In their study, they used knowledge tracing to examine how BKT parameters differed depending on whether students requested a hint. Similarly, in this analysis, we fit condition-specific BKT parameters to examine how spacing and color manipulations influence the latent learning processes captured in the model.

These comparisons are made possible by the \texttt{MultiBKT} model in StanBKT, with posterior inference performed via Markov chain Monte Carlo (MCMC). This approach contrasts with expectation-maximization (EM), as used in implementations such as \texttt{pyBKT}, which yields only point estimates and is inadequate for comparing parameters across the nine experimental conditions, particularly when some conditions contain relatively few observations. In contrast, MCMC produces full posterior distributions for each condition-specific parameter, allowing differences between conditions to be summarized with credible intervals and providing a principled assessment of uncertainty.

To summarize the effects of spacing and color in the cue conditions, we computed posterior differences for each condition-specific BKT parameter.
Specifically, for each posterior draw, we computed marginal parameter estimates by averaging over the levels of the complementary factor. Thus, spacing effects were obtained by averaging across the three color conditions within each spacing level (i.e., row-wise averages in Figure~\ref{fig:condition-design}), while color effects were obtained by averaging across the three spacing conditions within each color level (i.e., column-wise averages). For a generic BKT parameter $\phi \in \{P(T), P(F), P(G), P(S)\}$ the marginal effect of congruent spacing $\phi_{CS}$ was obtained by averaging the posterior draws for the three conditions with congruent spacing regardless of color condition \big(i.e., $\phi_{CS} = (\phi_{CSNC} + \phi_{CSCC} + \phi_{CSIC})/3$\big). Analogously, the incongruent and neutral spacing effects on the guess parameter, $\phi_{IS}$ and $\phi_{NS}$ respectively, were obtained by averaging over the corresponding sets of conditions. The same procedure was used to compute marginal effects for congruent $\phi_{CC}$, incongruent $\phi_{IC}$, and neutral $\phi_{NC}$ color cues.  

The effect of congruency in spacing and color was then defined as differences between each experimental condition and its neutral baseline. Specifically, the congruent spacing effect was computed as the difference between the marginal congruent spacing estimate and the marginal neutral spacing estimate (i.e., $\phi_{CS} - \phi_{NS}$), while the incongruent spacing effect compared incongruent spacing to neutral spacing (i.e., $\phi_{IS} - \phi_{NS}$). Similarly, the congruent and incongruent color effects were computed relative to the neutral color condition. 

For each contrast, we summarized the posterior distribution using the mean, median, and 95\% credible interval. Credible intervals that excluded zero were interpreted as evidence that the associated perceptual cue was related with a systematic change in the corresponding BKT parameter relative to the neutral condition.

The posterior contrasts are presented in Table~\ref{tab:spacing_color_contrasts}. For $P(T)$, no credible effects are observed, with all estimates centered near zero and credible intervals spanning zero across both spacing and color cues, indicating no meaningful influence on learning rate. Similarly, most effects for $P(F)$ and $P(S)$ are weak or uncertain. The only reliable deviations are a small reduction in the forgetting rate for the incongruent color condition ($-0.012$, 95\% CrI [$-0.023$, $-0.003$]) and a modest reduction in slip under the congruent spacing condition ($-0.030$, 95\% CrI [$-0.059$, $-0.001$]).

In contrast, the clearest and most consistent structure emerges in $P(G)$, where both spacing and color produce sizable and credible deviation from the neutral condition, particularly under incongruent conditions (spacing: $-0.058$ [$-0.093$, $-0.024$]; color: $-0.082$ [$-0.121$, $-0.040$]). These effects suggest reduced guessing under more structured or attention-demanding conditions.

\begin{table}[!htbp]
\centering
\begin{threeparttable}
\caption{Posterior contrasts for spacing and color effects on BKT parameters}
\label{tab:spacing_color_contrasts}
\begin{tabular}{lllcc}
\toprule
Parameter & Effect & Congruency & Mean Difference & 95\% CrI \\
\midrule

\multirow{4}{*}{Learn}
& \multirow{2}{*}{Spacing}
    & Congruent
    & -0.005
    & [-0.016, 0.008] \\
& & Incongruent
    & 0.002
    & [-0.010, 0.014] \\
& \multirow{2}{*}{Color}
    & Congruent
    & -0.004
    & [-0.017, 0.008] \\
& & Incongruent
    & 0.000
    & [-0.013, 0.012] \\

\midrule
\multirow{4}{*}{Forget}
& \multirow{2}{*}{Spacing}
    & Congruent
    & -0.000
    & [-0.010, 0.009] \\
& & Incongruent
    & 0.006
    & [-0.005, 0.016] \\
& \multirow{2}{*}{Color}
    & Congruent
    & -0.009
    & [-0.021, 0.002] \\
& & Incongruent
    & \textbf{-0.012}
    & \textbf{[-0.023, -0.003]} \\

\midrule
\multirow{4}{*}{Guess}
& \multirow{2}{*}{Spacing}
    & Congruent
    & 0.014
    & [-0.030, 0.055] \\
& & Incongruent
    & \textbf{-0.058}
    & \textbf{[-0.093, -0.024]} \\
& \multirow{2}{*}{Color}
    & Congruent
    & \textbf{0.046}
    & \textbf{[0.002, 0.091]} \\
& & Incongruent
    & \textbf{-0.082}
    & \textbf{[-0.121, -0.040]} \\

\midrule
\multirow{4}{*}{Slip}
& \multirow{2}{*}{Spacing}
    & Congruent
    & \textbf{-0.030}
    & \textbf{[-0.059, -0.001]} \\
& & Incongruent
    & -0.005
    & [-0.034, 0.025] \\
& \multirow{2}{*}{Color}
    & Congruent
    & -0.026
    & [-0.059, 0.007] \\
& & Incongruent
    & 0.031
    & [-0.002, 0.064] \\

\bottomrule
\end{tabular}
\tablenote{Values represent absolute differences relative to the neutral condition. Credible intervals excluding zero are shown in bold.
}
\end{threeparttable}
\end{table}

Overall, the results suggest that spacing and color cues are more closely associated with changes in $P(G)$ and, to a lesser extent, $P(S)$, rather than with learning or forgetting dynamics ($P(T)$, $P(F)$). Additionally, while incongruent conditions are associated with reduced guessing, consistent with increased attentional demand or more cautious responding, there is little evidence of changes in learning rates that indicate enhanced knowledge acquisition. Instead, the observed effects appear to be primarily related to students' responses rather than substantially altering underlying learning dynamics. These conclusions were facilitated by the Bayesian framework implemented in \textbf{StanBKT}, which was essential for the analysis by enabling full posterior inference over condition-specific BKT parameters and direct estimation of credible contrasts. This allowed uncertainty in parameter differences to be explicitly quantified, enabling more reliable differentiation between weak or ambiguous effects and those showing clearer shifts in model parameters.


\section{Discussion}

This paper introduced StanBKT, an open-source Python package that brings full Bayesian inference to Bayesian Knowledge Tracing using \texttt{Stan} as the computational backend. Although BKT has remained one of the most widely used and interpretable student modeling approaches, most existing software implementations are limited to point estimation via expectation-maximization. As a result, researchers typically obtain only a single set of parameter estimates, with no direct quantification of uncertainty in learning rates, forgetting rates, slip, guess, or initial mastery. StanBKT addresses this limitation by providing a unified and accessible framework for estimating BKT models using MCMC, variational inference, Pathfinder, and optimization-based methods.

In many applications, point estimates may be sufficient for prediction, and our empirical evaluation on the ASSISTments 2020 dataset demonstrated that predictive performance was effectively identical across EM, MAP, variational inference, Pathfinder, and MCMC. Thus, Bayesian inference should not be viewed as a means of improving predictive accuracy in classical BKT. Instead, its value lies in enabling richer statistical inference, principled regularization, and direct propagation of parameter uncertainty into downstream analyses.

This distinction is especially relevant in experimental and behavioral research, where the BKT parameters themselves are often the primary quantities of interest. Parameters such as the learning and forgetting rates are commonly interpreted as indicators of long-term knowledge acquisition and retention, while slip and guess rates are used to characterize immediate performance and scaffolding effects \parencite{beckDoesHelpHelp2008, sao2013incorporating}. In these settings, uncertainty quantification is critical for assessing whether observed differences between conditions are substantively meaningful. StanBKT facilitates this type of analysis by providing full posterior distributions, credible intervals, and posterior predictive summaries, allowing researchers to make probabilistic statements about parameter differences rather than relying solely on point estimates.

StanBKT also provides a flexible framework for incorporating prior information and extending BKT to more complex modeling structures. Priors can be used to encode substantive knowledge, regularize estimation, and mitigate well-known identifiability issues in BKT. The package supports standard, grouped, and hierarchical variants, enabling analyses that range from fully pooled models to partially pooled models that borrow strength across groups. This is particularly valuable in educational experiments and small-sample studies, where some groups may contain relatively few observations and independent estimation can lead to unstable parameter estimates.

The worked example and empirical evaluations illustrate these capabilities. In the simulated example, StanBKT recovered the true data-generating parameters and provided standard posterior diagnostics, demonstrating the transparency and interpretability of the Bayesian workflow. In the ASSISTments benchmark, the results showed that approximate inference methods such as variational inference and Pathfinder achieved predictive performance equivalent to MCMC while requiring substantially less computation. These findings suggest that researchers can choose an inference method that matches their analytic goals: MCMC when accurate uncertainty quantification is required, and faster approximate methods when scalability is a greater concern.

StanBKT is particularly well suited for several use cases. First, it provides a convenient package for methodological comparisons across inference algorithms and prior specifications. Second, it enables experimental researchers to estimate and compare BKT parameters across intervention conditions while accounting for uncertainty. Third, the hierarchical formulation offers a principled approach for partially pooled estimation in sparse-data settings. Finally, the package integrates modern Bayesian diagnostics and visualization tools into an interpretable student modeling framework, making it useful both for applied educational research and methodological development.

Several limitations should be acknowledged. Bayesian inference via MCMC remains substantially more computationally and memory intensive than EM-based implementations such as \texttt{pyBKT}. This cost scales with both the number of chains and posterior draws, as each chain stores full parameter trajectories and intermediate states, which can become memory-intensive in large-scale or high-granularity analyses. Although StanBKT supports faster alternatives such as MAP estimation, variational inference, and Pathfinder, these approaches reduce computational burden but introduce additional approximation error and may underestimate posterior uncertainty, particularly in complex or weakly identified models.

Despite these limitations, StanBKT substantially lowers the barrier to using Bayesian inference for knowledge tracing. By combining the interpretability of classical BKT with the flexibility of probabilistic programming, the package enables researchers to move beyond point estimation and treat uncertainty as a central component of inference. This is particularly important when BKT is used not only to predict student performance, but also to test educational theories, compare interventions, and draw substantive conclusions about learning processes.

Overall, StanBKT provides a flexible and extensible framework for estimating Bayesian Knowledge Tracing models using modern inference techniques. The package preserves the conceptual simplicity and interpretability that have made BKT a foundational model in educational data mining, while adding principled uncertainty quantification, hierarchical modeling, and support for multiple inference algorithms. For researchers seeking an interpretable student model that supports rigorous statistical inference, StanBKT offers a practical and powerful alternative to traditional EM-based implementations.

\section{Acknowledgments}
We gratefully acknowledge the Stan Development Team for their continued development and maintenance of the Stan probabilistic programming language, which provides the core inference engine underlying StanBKT. This material is based upon work supported by the National Science Foundation under Grant [2300764] awarded to Worcester Polytechnic Institute.

\section{Declaration}

\subsection{Funding}
This material is based upon work supported by the National Science Foundation under Grant Number [2300764] to Worcester Polytechnic Institute. Any opinions, findings, and conclusions or recommendations expressed in this material are those of the author(s) and do not necessarily reflect the views of the National Science Foundation.

\subsection{Conflicts of interest}
The authors have no relevant financial or non-financial interests to disclose.

\subsection{Conflicts of interest}
The authors have no relevant financial or non-financial interests to disclose.

\subsection{Ethics approval}
Not applicable.

\subsection{Consent to participate}
Not applicable.

\subsection{Consent for publication}
Not applicable.

\subsection{Availability of Data and Materials}

The source code for the StanBKT package is publicly available on GitHub at
\href{https://github.com/SiddharthaPradhan/StanBKT}{github.com/SiddharthaPradhan/StanBKT}.

The simulated data and analysis scripts used in the
\nameref{subsec:worked-example} and the two experiments reported in
\nameref{sec:evaluation} are available at
\href{https://github.com/SiddharthaPradhan/StanBKT-Experiments}{github.com/SiddharthaPradhan/StanBKT-Experiments}.

The data used for the runtime evaluation described in
\nameref{subsec:assist-eval} are available on the Open Science Framework (OSF) at
\href{https://osf.io/q7zc5}{osf.io/q7zc5}. The data used for the cue condition
comparison described in \nameref{subsec:cue-eval} are available at
\href{https://osf.io/d3tzq}{osf.io/d3tzq}.

\subsubsection{Authors' contributions}
Siddhartha Pradhan: Conceptualization, Package Development, Software Engineering, Methodology, Analysis, Writing – Original Draft, Writing – Review \& Editing, Supervision. Yanping Pei: Writing – Review \& Editing. Morgan Lee: Writing – Review \& Editing. Puyuan Zhang: Writing – Review \& Editing. Erin Ottmar: Conceptualization, Writing – Review \& Editing, Funding Acquisition, Supervision, Project Administration.  Adam Sales: Conceptualization, Methodology, Analysis, Writing – Review \& Editing, Supervision.

\printbibliography

\end{document}